\documentclass[amsmath,amssymb,twocolumn,nofootinbib,floatfix,superscriptaddress]{revtex4-1}
\usepackage{float}
\usepackage{soul}
\usepackage{amsmath}
\usepackage{graphicx}
\usepackage{dcolumn}
\usepackage{bm}
\usepackage{natbib}
\usepackage{tikz}
\usepackage{color}
\usepackage{subfigure}
\usepackage{longtable}
\usepackage{dcolumn}

\usepackage[colorlinks, linkcolor={blue},citecolor={blue},breaklinks]{hyperref}
\begin{document}

\title{Production of Quasi-Stellar Neutron Field at Explosive Stellar Temperatures}    

\author{M.~Friedman}
\email{moshe.friedman@mail.huji.ac.il}
\affiliation{Racah Institute of Physics, Hebrew University, Jerusalem, Israel 91904}

\date{\today}

\begin{abstract}
Neutron-induced reactions on unstable isotopes play a key role in the nucleosynthesis $i$--, $r$--, $p$--, $rp$-- and $\nu p$--processes occurring in astrophysical scenarios. While direct cross section measurements are possible for long-living unstable isotopes using the neutron Time-of-Flight method, the currently available neutron intensities ($\approx10^{6}$ n/s) require large samples which are not feasible for shorter lifetime isotopes. For the last four decades, the $^{7}$Li$(p,n)$ reaction has been used to provide a neutron field at a stellar temperature of $\approx$ 0.3 GK with significantly higher intensity, allowing the successful measurement of many cross sections along the $s$-process path. In this paper we describe a novel method to use this reaction to produce neutron fields at temperatures of $\approx$ 1.5-3.5 GK, relevant to scenarios such as convective shell C/Ne burning, explosive Ne/C burning, and core-collapse supernovae. This method will allow direct cross section measurements of many important reactions at explosive temperatures, such as $^{26}$Al$(n,p)$, $^{75}$Se$(n,p)$ and $^{56}$Ni$(n,p)$.

\end{abstract}

\maketitle

\section{Introduction}
\label{intro}
Characterization of neutron-induced reaction rates on unstable isotopes is essential for a complete understanding of the nucleosynthesis occurring in many astrophysical sites and events. This information is used as input for network calculations that provide predictions on the expected abundances \cite{Wiescher12,Frohlich14,Rauscher14,Jose16}. Comparing these predictions to the observed abundances, may, for example, provide an indication either on the accuracy of the models, the contribution of different astrophysical events to the galactic isotopic content, or the frequency of specific events such as supernovae.

Direct cross section measurements require the production of a sample containing a sufficient quantity of the desired unstable isotope, a neutron field with a well-defined energy distribution at relevant energies for astrophysics (typically 1-2000 keV), and a high neutron intensity. These requirements impose significant limitations on the feasibly-accessible isotopes for direct cross section measurements. Production of large quantities of unstable isotopes involves challenges regarding hot chemistry, radiation protection and disposal. In some cases highly-active samples induce high background which makes the detection of the desired quantities more difficult. These challenges are strongly correlated with the lifetime of the studied isotopes, making short-living isotopes extremely hard to study.

The need for a well-defined neutron energy distribution makes the neutron Time-of-Flight (TOF) technique the most straight-forward approach for such experiments. In this technique, a white, pulsed, neutron source (e.g., via spallation) is located at a fair distance from the sample, and the neutron energy is calculated on an event-by-event basis using the known flight distance and the time difference between neutron production and detection \cite{Reif2018}. The energy resolution of such systems is mainly determined by the time resolution of the pulse and the distance between the source and the sample. Consequently, TOF facilities use long flight distances, typically tens of meters and above, resulting in a dramatic reduction of the neutron intensity. The CERN n\textunderscore TOF facility EAR2, currently the most intense neutron TOF facility, provides neutron intensities on the order of $10^6$ neutrons per second at the relevant energy range for astrophysics \cite{EAR2}. An alternative approach suggested by Ratynski and k{\"a}ppeler \cite{kappler88}, has been since extensively used for the last four decades \cite{kappler11}. They suggested to use the $^7$Li$(p,n)$ reaction as a neutron source, with a specific proton energy of $E_p=1912$ keV, close to the reaction threshold of $E_p=1880$ keV, and a thick Li target. At this proton energy, the neutrons are forward-collimated, and their energy distribution resembles that of Maxwellian flux distribution with temperature of $\approx 0.3$ GK, allowing nearly-direct measurement of the Maxwellian-Averaged Cross-Section (MACS) at this temperature, which is relevant for some $s$-process sites \cite{kappler11}. An important advantage of this approach is a substantial increase in the neutron intensity, up to $\approx 10^9$ neutrons per second for 100--$\mu$A proton beam. Recently, the intense proton beam of SARAF (up to $\approx$ 2 mA at the relevant) allowed measurements with an even higher neutron intensity on the order of $10^{10}$ neutrons per second \cite{LiLiTreview}. The 4 orders of magnitude increase in the neutron intensity allows measurements of smaller cross section reactions or shorter-lived isotopes. 

In spite of the successful use of the $^7$Li$(p,n)$ reaction close to threshold energy as an intense neutron source, it is limited for a specific temperature of $\approx 0.3$ GK. Suggestions were made to similarly use the $^{18}$O$(p,n)$ reaction to produce a neutron field at $\approx 0.06$ GK \cite{Heil2005}, and to use a varying proton beam energy on the lithium target to achieve an effective neutron field at $\approx 1$ GK \cite{Reif2018}. Reifarth \textit{et al.} \cite{Reif2008} also demonstrated the use of a higher proton beam energy on a thin Li target to produce higher-energy neutrons. In this paper we describe a novel approach to extend the use of a varying proton beam energy on thick and thin lithium targets to study neutron-induced cross sections in the temperature region of 1.5-3.5 GK. The method allows the study of neutron-induced reactions continuously throughout the temperature region, rather than at a specific temperature. Particularly, we will discuss the use of this technique for direct measurements of $(n,p)$ and $(n,\alpha)$ cross sections at explosive stellar temperatures, a region where experimental data is poor or even non-existent for many important nuclei.

In Sec. \ref{sec:method} we will describe the method to produce the neutron field and the required computational tools, in Sec. \ref{sec:uses} we will discuss specific optional  applications for the method, and we will summarize in Sec. \ref{sec:summary}.

\section{Method}
\label{sec:method}
\subsection{General Scope}
\label{sec:general}
The goal is to use the $^7$Li$(p,n)$ reaction at varying proton energies for direct measurements of neutron-induced reactions on short-lived isotopes with neutron energy spectra that resemble a Maxwellian flux with any temperature at the range of 1.5-3.5 GK. The assumption is that experimental constrains, related to the activity of the sample and the high-radiation environment, will require flexibility in the geometry of the experiment, primarily in terms of sample position and shielding. Our main interest are $(n,p)$ reactions on proton-rich nuclei, which determines the requirement for prompt measurements, as the products will be dominated by spontaneous $\beta^+$ decay. The experimentalist will need the proper tools to plan the specific experiment within his constrains, and to extract the temperature-dependent cross section. 

\subsection{Obtaining MACS}
The astrophysical reaction rate for neutron induced reactions on isotope $X$ is given by \cite{Beer1992}:
\begin{equation}
  r = N_nN_X\left<\sigma v\right>,
\end{equation}
where $N_n$ is the neutron density, $N_X$ is the density of isotope $X$, $\sigma$ is the energy-dependent cross-section, $v$ is the relative velocity between the particles, and $\left<\sigma v\right>$ is:
\begin{equation}
  \left<\sigma v\right>  = \int{\sigma(v)v\phi(v)dv}
\end{equation}
The velocity distribution follows the Maxwell-Blotzmann distribution $\phi_{MB}(v)$ with the appropriate temperature. It is customary to define the Maxwellian-Averaged Cross-Section (MACS):
\begin{equation}
   v_T = \sqrt{\frac{2\overline{E}}{m}}
\end{equation}
\begin{equation}
  \textrm{MACS} = \frac{\left<\sigma v\right>_{MB}}{v_T}
\end{equation}
Which, in terms of energy, is:
\begin{equation}
  \textrm{MACS} = \frac{2}{\sqrt{\pi}}\frac{1}{(kT)^2}\cdot\int{\sigma(E)\cdot E\cdot e^{-\frac{E}{kT}}}\cdot dE. \label{eq:MACS}
\end{equation}
In the lab, the target nuclei is at rest, and for a given experimental neutron energy distribution $\phi_{exp}=(dn/dE)_{exp}$ the measured cross section will be:
\begin{equation}
  \sigma_{exp} = \int{\sigma({E})\cdot\phi_{exp}\cdot dE}. \label{eq:sigma_exp}
\end{equation}
Our goal is to obtain $\phi_{exp}$ as close as possible to $\phi_{MB}$:
\begin{equation}
  \phi_{MB} = \frac{1}{(kT)^2}E\cdot e^{-\frac{E}{kT}}    \label{eq:MBflux}
\end{equation}
for the extraction of the MACS values. While this is possible for $kT\sim30$ keV, increasing the proton energy, although results in higher neutron energy, does not produce similar neutron-energy distribution. Instead, we propose to perform a set of $I$ measurements at different proton energies $E_p^i$. For each proton energy, a corresponding neutron energy distribution $\phi_{exp}^i$ is obtained, and a cross section $\sigma_{exp}^i$ is extracted. If we know the neutron energy distribution $\phi_{exp}^i$, we can try to find a set of weights $W^i$ that will satisfy:
\begin{equation}
  \overline{\phi_{exp}} = \frac{1}{\sum{W^i}}\cdot \sum{W^i\cdot\phi_{exp}^i} = \phi_{MB}. \label{eq:superposition}
\end{equation}
If we can indeed find such a set of $W^i$, we can obtain a weighted-averaged experimental cross section:
\begin{equation}
  \overline{\sigma_{exp}} = \frac{1}{\sum{W^i}}\cdot \sum{W^i\cdot\sigma_{exp}^i} \propto \textrm{MACS}. 
\end{equation}
Practically, such a procedure will require a large number of measurements $I$ with a very thin Li target that will produce $\phi_{exp}^i$ which are closely mono-energetic (or with two sharp energy peaks for $E_p > 2.373$ MeV), at the cost of a lower neutron intensity. Instead, it is possible to use a thicker Li target and search for a set of $W^i$ that will give a good approximation of Eq. \ref{eq:superposition}. The exact lithium target thickness and proton beam energies should be determined for each specific experiment as will be demonstrated in the examples in Sec. \ref{sec:uses}. It should be noted that the same set of measurements to obtain $\sigma_{exp}^i$ can be used with different weights to obtain different temperatures. The required computational tools for this task are discussed below.

\subsection{Optimizing Experimental Parameters}
\begin{figure}
  \centering
    \includegraphics[width=\linewidth]{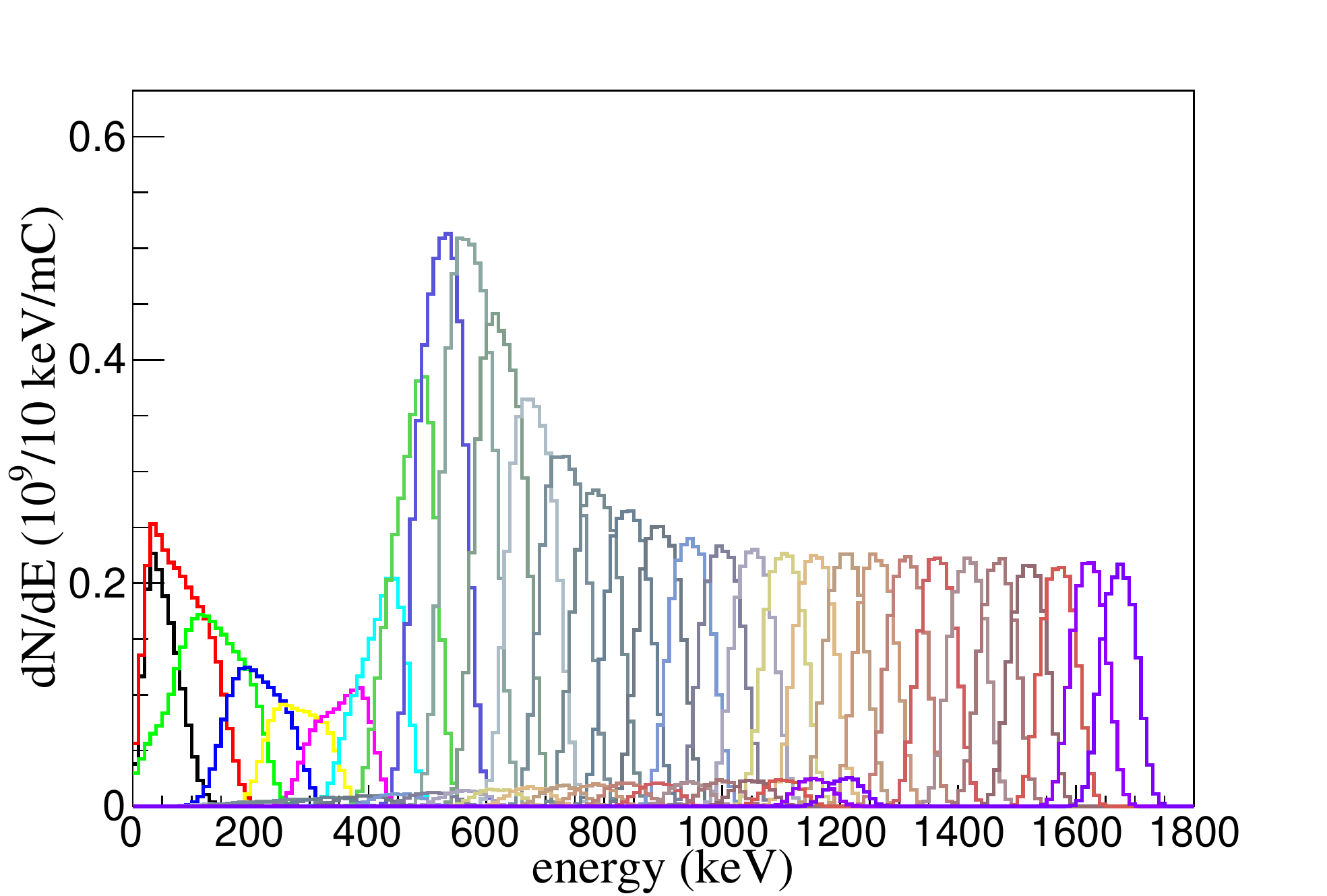}
    \includegraphics[width=\linewidth]{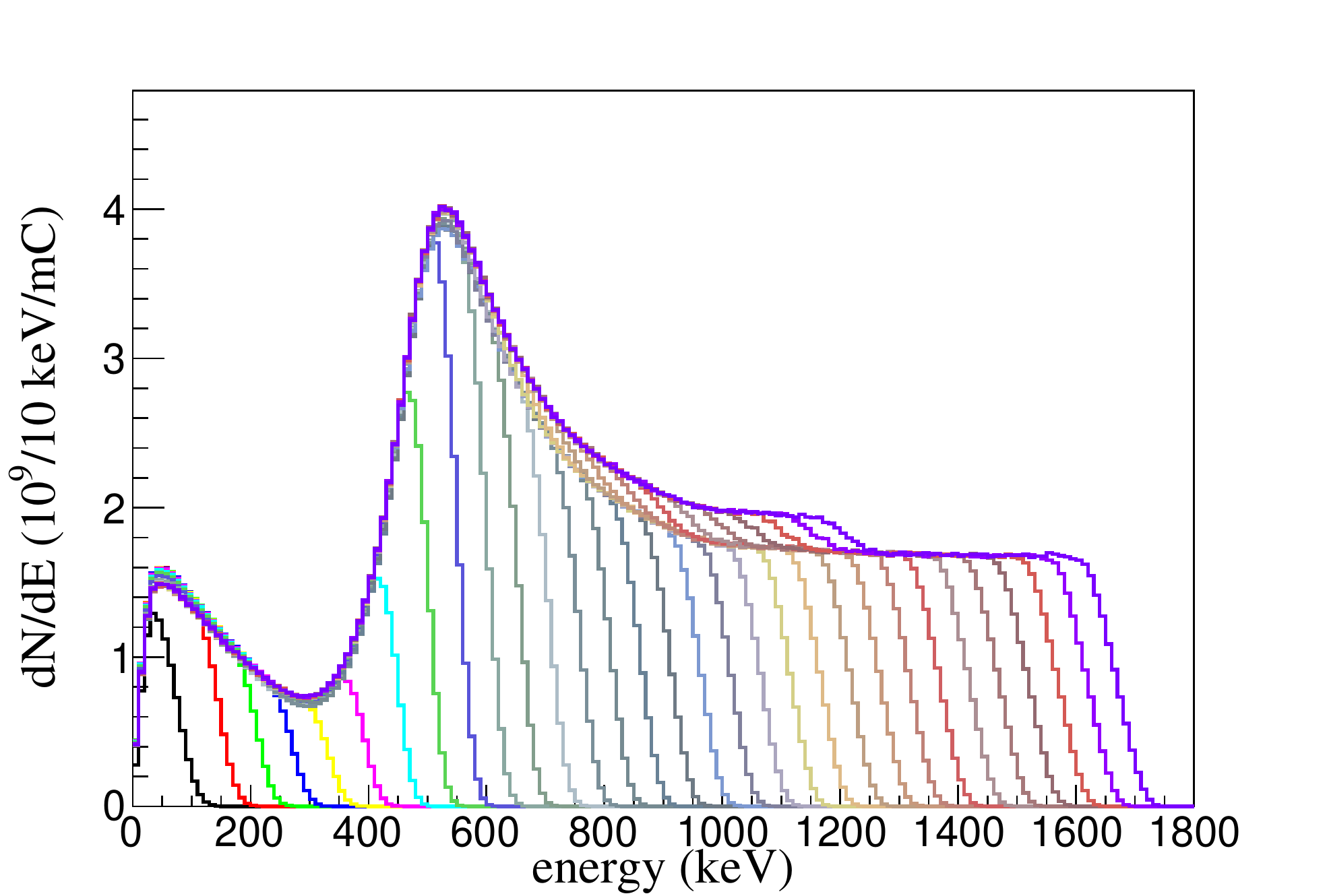}
  \caption{SimLiT calculated neutron spectra for different proton energies. The different colors represents different proton energies from 1900--keV to 3400--keV in 50--keV intervals. The top panel is calculated for a 15 $\mu$m-thick lithium target, and the sample is a  5--cm\textsuperscript{2} disk, 7 cm downstream the lithium target. The bottom panel is calculated for a thick lithium target, and the sample is a 1--cm\textsuperscript{2} disk, 1 cm downstream the lithium target.}
  \label{fig:base}
\end{figure}

Generally, protons impinging on a thin lithium target and a sample positioned so that it covers a small solid angle, will result in a quasi-mono-energetic neutron spectrum and a low neutron intensity, while a thick lithium target and a large solid angle will result in a wider neutron energy distribution and a higher neutron intensity. A simulation of the neutron spectrum and intensity at the sample is required for optimizing the experimental assembly. Several Monte-Carlo simulation codes for the $^7$Li$(p,n)$ neutron yields exist \cite{PINO,simlit,Herrera2015,Pachuau2017}. In this paper we use the SimLiT code \cite{simlit}. The SimLiT code was compared to experimental data at $E_p = 1912$ keV and $E_p = 2000$ keV \cite{simlit}, and to other simulations at higher energies \cite{Pachuau2017}. It was also successfully used for many experiments \cite{LiLiTreview}. Fig. \ref{fig:base} shows SimLiT calculated neutron spectra for proton energies of 1900--keV to 3400--keV. The figure compares two scenarios. In the first scenario (Fig. \ref{fig:base} top) the lithium target thickness is 15 $\mu$m, and the sample is a 5--cm\textsuperscript{2} disk, 7 cm downstream the lithium target. Evidently, the resulting neutron spectra are relatively convenient for finding a set of $W^i$ that will satisfy Eq. \ref{eq:superposition}, and the neutron intensities vary between $1.5\times10^9$ and $5\times10^9$ n/s/mC for the different proton energies. Assuming a 100--$\mu$A proton beam, this translates to $\sim10^8$ n/s on the sample - two orders of magnitude improvement compared to the EAR2 neutron TOF facility \cite{EAR2}. In the second scenario (Fig. \ref{fig:base} bottom) the lithium target is thick enough to slow the proton beam below the reaction threshold. The sample is a 1--cm\textsuperscript{2} disk, 1 cm downstream the lithium target. In this case, the neutron spectra are smeared all over the allowed energy region, with a unique structure at the energy region between 100-- and 600--keV. As will be shown below, it is harder to find a set of $W^i$ that will satisfy Eq. \ref{eq:superposition}, and one needs to find the best possible approximation of it. On the other hand, the neutron intensities vary between $1\times10^{10}$ and $3\times10^{11}$ n/s/mC for the different proton energies. Furthermore, thick lithium target can withhold much higher proton beam intensities, and 1-2 mA beams are already available \cite{LiLiTreview}. This allows a total neutron flux of $\sim10^{11}$ n/s/cm$^2$ on the sample - five orders of magnitude improvement compared to the EAR2 neutron TOF facility. Future neutron-production facilities \cite{FRANZ} will enable even higher neutron intensities.

\begin{figure}
  \centering
    \includegraphics[width=\linewidth]{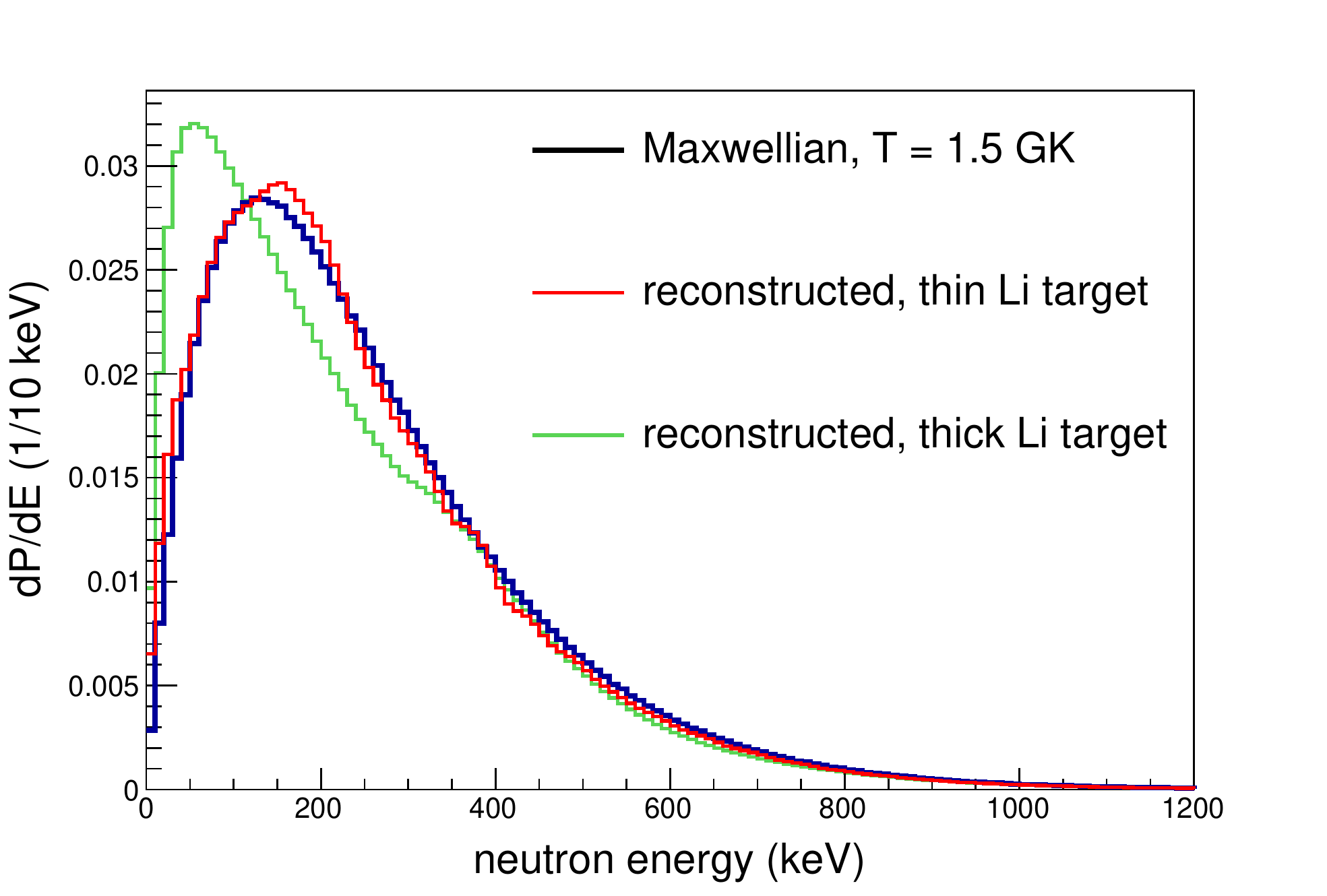}
    \includegraphics[width=\linewidth]{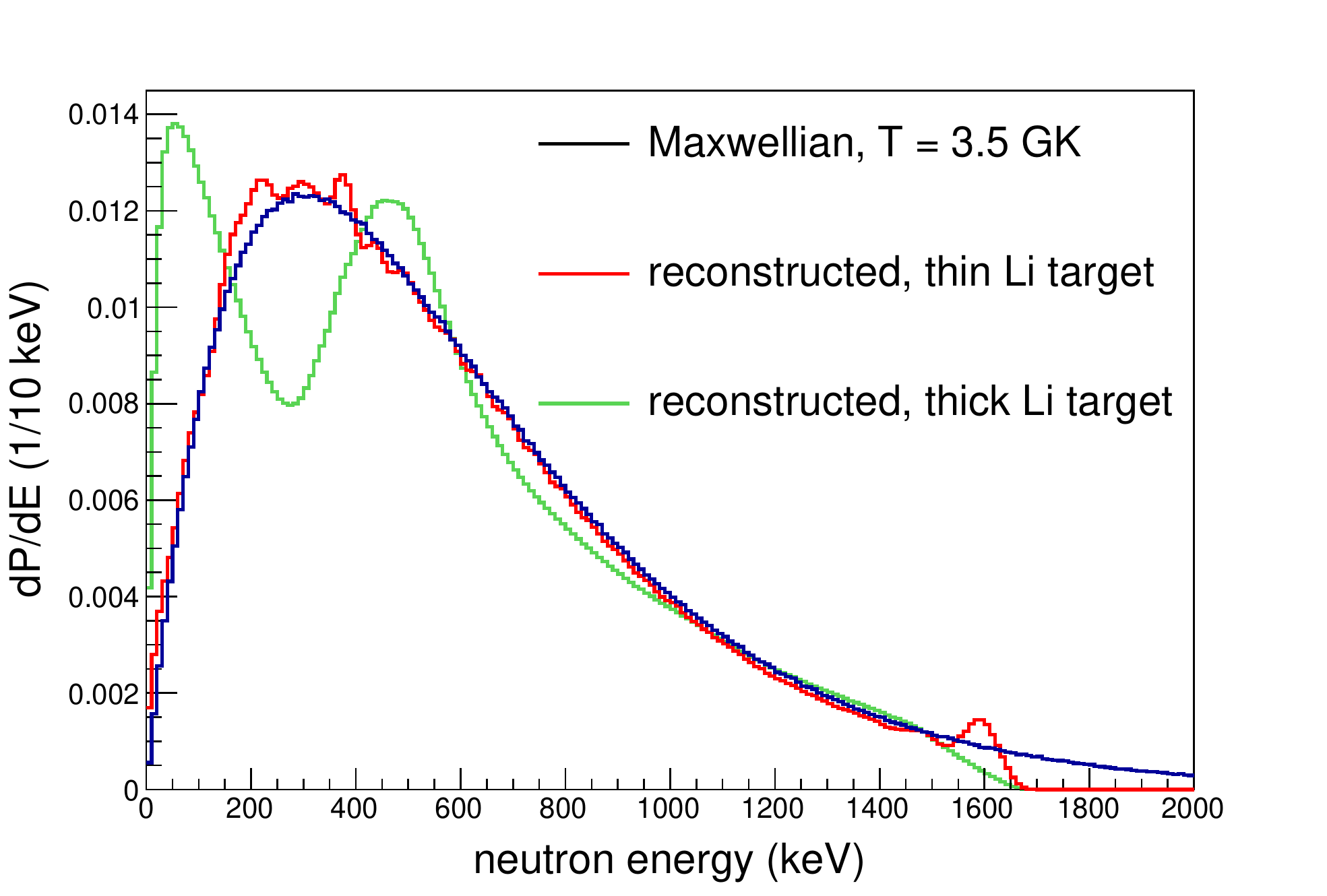}
  \caption{Reconstructed Maxwellian flux distribution  using the basis distribution functions obtained by thin (15 $\mu$m) and thick lithium targets (See Fig. \ref{fig:base}). Top pannel shows the best reconstruction for a temperature of 1.5 GK, while the bottom pannel shows the best reconstruction for a temperature of 3.5 GK using the same basis functions.}
  \label{fig:reconstructed}
\end{figure}

For the purpose of this paper, the weighting parameters $W^i$ were calculated using the RooUnfold package \cite{RooUnfold}, specifically, we found the Bayesian Unfolding method \cite{bayesUnfold} within this package to be the most successful one. The input for the unfolding process are the individual-energy spectra (basis functions), presented in Fig. \ref{fig:base}, and a goal function, in this case Maxwellian flux distribution. The ouput is a set of weights $W^i$ for the basis functions that will most closely reconstruct the goal function. Fig. \ref{fig:reconstructed} shows two reconstructed spectra for the two set of basis function presented in Fig. \ref{fig:base}. The goal functions presented are Maxwellian flux distributions for temperatures of 1.5 and 3.5 GK. As can be seen in Fig. \ref{fig:reconstructed}, Eq. \ref{eq:superposition} is closely satisfied with the thin lithium target basis, whereas in the case of the thick lithium target the low-energy part is not reconstructed as well. The effect of the differences on the obtained MACS should be studied in the context of the specific measurement, as demonstrated in Sec. \ref{sec:uses}. 

\section{Specific Application Examples}
\label{sec:uses}
To demonstrate the use of the discussed method, we chose three important reactions for explosive nucleosynthesis: $^{26}$Al$(n,p)$, $^{75}$Se$(n,p)$ and $^{56}$Ni$(n,p)$.

\subsection{$^{26}$Al$(n,p)$ and $^{26}$Al$(n,\alpha)$}
\label{sec:al26}
\begin{figure}
  \centering
    \includegraphics[width=\linewidth]{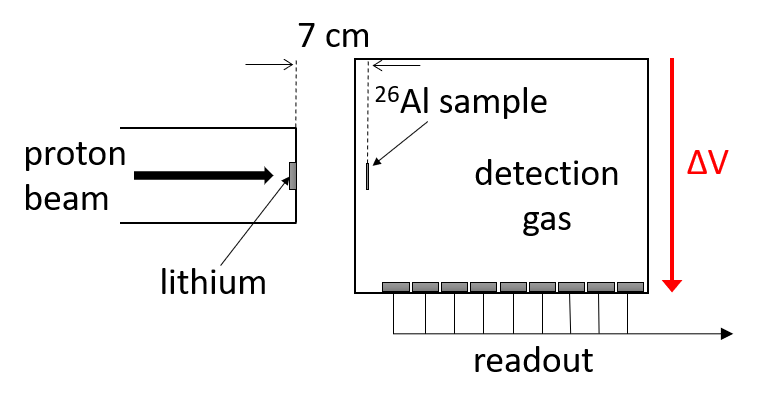}
  \caption{Conceptual setup for a measurement of the $^{26}$Al$(n,p)$ cross section. The proton beam is impinging on a 15--$\mu$m-thick lithium target to produce neutrons. The 5--cm$^{2}$ $^{26}$Al sample is placed 7 cm downstream the lithium target in a gas-filled chamber which is held under a potential gradient. The protons are detected with 50\% efficiency by ionizing electrons in the gas. The ionization electrons are drifted under a constant electric field and collected by a segmented readout plane for particle identification purposes.}
  \label{fig:al26}
\end{figure}

Constraining destruction rates of $^{26}$Al ($t_{1/2} = 7.2\times10^5$ y) via $(n,p)$ (and $(n,\alpha)$) in convective shell C/Ne Burning and explosive Ne/C burning is expected to dominate the uncertainties of $^{26}$Al production in those sites \cite{Iliadis2011}, which is an important observable for $\gamma$-ray astronomy \cite{Diehl2006}. Even though the importance of these rates is known for a long time, only few measurements were performed at stellar energies \cite{Trautvetter86,Koehler97,DeSmet07}, of which only a single measurement was performed at the relevant energies for explosive scenarios, and it only covered a portion of the relevant energy region: 270-350 keV \cite{Trautvetter86}, hence these rates still require investigation at explosive stellar temperatures \cite{Wiescher12,Iliadis2011}. The suggested method in this paper can be used to measure this reaction using available facilities. 

Fig. \ref{fig:al26} shows a conceptual assembly of the experiment. A 15--$\mu$m--thick lithium target is used to produce the neutron spectra described in Fig. \ref{fig:base} top pannel. A gas filled detector is positioned $\sim$5 cm downstream the lithium target. A $^{26}$Al sample of $1.2\times10^{17}$ atoms (0.1 $\mu$Ci) is deposited on a 5--cm$^2$ $^{27}$Al backing, and positioned 7 cm downstream the lithium target. Aluminum backing is chosen due to the high thershold for $^{27}$Al$(n,p)$ and $(n,\alpha)$ reactions (1828--keV and 3133--keV, respectively). The protons and alphas are emitted in all directions, which reduces the detection efficiency to 50\%. The chamber is held under a constant electric field. The charged particles are detected by ionizing the gas, and the ionization electrons are drifted towards a segmented readout plane. The segmentation provides particle identification using the $\Delta E-E$ technique, to distinguish between protons and alphas that are emitted from the sample, and possible ionization from other sources, such as secondary activation on the detector walls by neutrons.

Detection rates vary with the varying proton energies, described in Sec. \ref{sec:method}. Assuming a 10--$\mu$A proton beam current, we estimate 100-500 $(n,p)$ and 30-150 $(n,\alpha)$ detections per hour of irradiation. This estimation relies on the TENDL-2017 cross section library \cite{TENDL}.

As discussed in Sec. \ref{sec:method}, the reconstructed neutron spectrum does not perfectly match the Maxwellian flux distribution, hence Eq. \ref{eq:superposition} is not perfectly satisfied. There is no \textit{a priori} way to calculate the error introduced by the differences, but we can make a rough estimation based on an existing cross section library. We used the TENDL-2017 cross section library throughout this paper. Table \ref{tab:errors} shows the differences between the averaged cross sections calculated using a Maxwellian flux with temperatures of 1.5, 2.5 and 3.5 GK, and the corresponding calculated cross sections using the reconstructed spectra. From the table it is evident that, based on TENDL-2017, the two neutron spectra are practically identical for this specific measurements. However, this estimation does not consider the possibility for unknown resonances that might increase the errors. 

\begin{table*}
\centering
\begin{ruledtabular}
  \begin{tabular}{lcccccc}
   & \multicolumn{6}{c}{calculated cross section (mb)} \\
   & \multicolumn{2}{c}{T = 1.5 GK} & \multicolumn{2}{c}{T = 2.5 GK} & \multicolumn{2}{c}{T = 3.5 GK} \\
   reaction & Maxwellian & reconstructed & Maxwellian & reconstructed & Maxwellian & reconstructed \\   
   \hline
   \\
   $^{26}$Al$(n,p)$      & 266  & 265  & 250  & 249  & 242  & 241 \\
   $^{26}$Al$(n,\alpha)$ & 61.2 & 60.5 & 65.9 & 65.0 & 73.3 & 71.3\\
   $^{56}$Ni$(n,p)$      & 223  & 213  & 296  & 279  & 367  & 341 \\
   $^{75}$Se$(n,p)$      & 1.15 & 1.09 & 1.82 & 1.68 & 2.59 & 2.31\\ 
  \end{tabular}
  \label{tab:errors}
  \caption{TENDL-2017 cross sections, averaged over Maxwellian flux distributions and the reconstructed distributions (see Fig. \ref{fig:reconstructed}), for the reactions discussed in Sec. \ref{sec:uses}. All calculations were done using TENDL-2017 cross sections library \cite{TENDL}. Cross sections are given in mb.}
\end{ruledtabular}
\end{table*}

\subsection{$^{56}$Ni$(n,p)$ and $^{75}$Se$(n,p)$}
\label{sec:ni56}
Constraining the rate of the $^{56}$Ni$(n,p)$ reaction at temperatures of 1.5-3 GK was identified to be important for studies of the $\nu p$--process in core-collapse supernova \cite{Frohlich14,Rauscher14}. The short lifetime of $^{56}$Ni ($t_{1/2} = 6.075$ d) makes a direct measurement of this reaction extremely challenging, and currently there is no experimental data available. Efforts were made to constrain this rate using the inverse $^{56}$Co$(p,n)$ reaction \cite{Gastis16}, and directly with the neutron-TOF technique \cite{Kuvin19}. Similar difficulties are expected for measuring the $^{75}$Se$(n,p)$ cross section ($t_{1/2} = 119.8$ d), which was identified as important for the $p$-process \cite{Rapp06}.

\begin{figure}
  \centering
    \includegraphics[width=\linewidth]{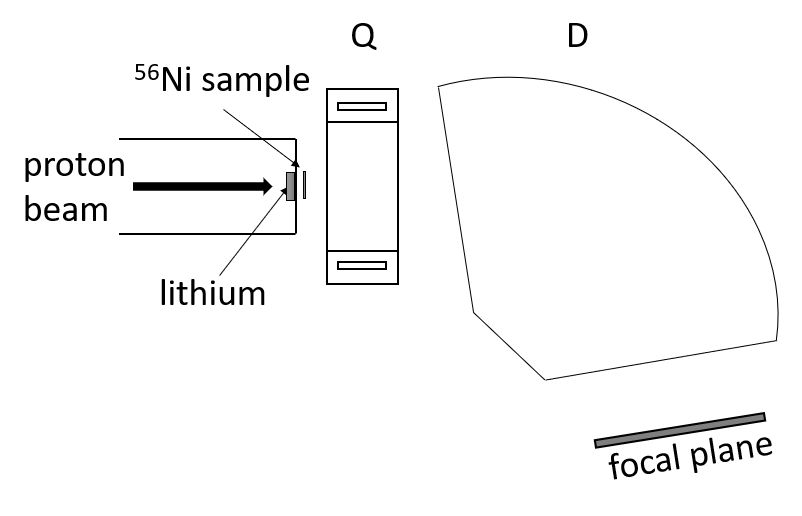}
  \caption{Conceptual setup for a measurement of $^{56}$Ni and $^{75}$Se $(n,p)$ cross sections. The proton beam is impinging on a thick lithium target to produce neutrons. The 1--cm$^2$ sample is placed 1 cm downstream the lithium target in a vacuum environment, and directed through a quadrupole-dipole (QD) magnet configuration to be detected at a focal plane $\sim$3.5 m from the sample. }
  \label{fig:ni56}
\end{figure}

The short lifetimes of these isotopes impose a limitation on the sample size, from both production and handling perspectives. The advantage of the method discussed in this paper is the much higher neutron intensity, which allows the use of smaller samples. However, using low beam intensity and thin lithium target as in Sec. \ref{sec:al26} is not sufficient, and the high neutron, $\beta^+$, and $\gamma$ radiation from the target itself and from the more intense beam does not allow to place the detector in such close proximity to the sample. Instead, we suggest to use a thick lithium target and as high beam intensity as possible. For detection, one can use a dipole magnet to direct the emitted protons towards a detector at a focal plane at some distance from the sample (see Fig. \ref{fig:ni56}). A tentative design of a QD spectrometer suggests about 1\% acceptance for such an assembly \cite{LASI}.

Target production will be possible in the isotope-harvesting facility, currently under construction at FRIB \cite{Abel19}. For this discussion, we assume a conservative sample size of 1 mCi ($2.8\times10^{13}$ and $5.6\times10^{14}$ atoms for $^{56}$Ni and $^{75}$Se samples, respectively). Assuming a 1 mA proton beam, and considering 1\% acceptance of the spectrograph, we estimate the detection rate to be 1-150 and 0.1-26 detections per hour of irradiation for the $^{56}$Ni and $^{75}$Se experiments, respectively. This might suggest that a larger sample size will be recommended for the $^{75}$Se case.

The use of a thick lithium target produces a broader neutron spectra as discussed in Sec. \ref{sec:method} and illustrated in Fig. \ref{fig:base} (bottom), which results in a reconstructed spectrum less similar to the Maxwellian flux distribution. The effect of the differences was studied similarly to the $^{26}$Al case and estimated to be at the 10\% level (see Table \ref{tab:errors}).

\section{Summary}
\label{sec:summary}
Measurements of neutron-induced cross sections on unstable nuclei at explosive stellar temperatures require high-intensity neutron fluxes at the relevant energies. We studied the possibility to use the $^7$Li$(p,n)$ reaction with varying proton energies to produce a set of cross section measurements that can be combined with proper weights to obtain MACS values at temperatures of 1.5-3.5 GK. This approach provides up to 5 orders of magnitude improvement in the neutron intensity over the state-of-the-art neutron-TOF facilities. We demonstrated the advantage of this technique for measurements of important $(n,p)$ reactions that are not easily accessible with the neutron-TOF method.

\section*{Acknowledgements}
The authors would like to thank Prof. G. Berg from Notre Dame University and Dr. E. Pollacco from CEA-Saclay for providing advice regarding ion transport and detection techniques. We would also like to thank Dr. M. Tessler from SNRC-SARAF for consultation.


\begin{thebibliography}{100}

\bibitem{Wiescher12}M. Wiescher, 1 F. K{\"a}ppeler, 2 and K. Langanke, Annu. Rev. Astron. Astrophys. 50:165–210 (2012).
\bibitem{Frohlich14}C. Fr{\"o}hlich, J. Phys. G: Nucl. Part. Phys. 41 044003 (2014).
\bibitem{Rauscher14}T. Rauscher, AIP Advances 4, 041012 (2014).
\bibitem{Jose16}J. Jos{\'e}, Stellar Explosions: Hydrodynamics and Nucleosynthesis, CRC Press, Boca Raton, (2016).
\bibitem{Reif2018}R. Reifarth \textit{et al.}, Eur. Phys. J.  Plus, \textbf{133}, 424 (2018).
\bibitem{EAR2}C. Wei{\ss} \textit{et al.}, Nucl. Instrum. Method. A, 799 0168-9002 (2015).
\bibitem{kappler88} W. Ratynski, F. K{\"a}ppeler, Phys. Rev. C 37 595 (1988).
\bibitem{kappler11}F. K{\"a}ppeler, R. Gallino, S. Bisterzo, and Wako Aoki, Rev. Mod. Phys. 83, 157 (2011).
\bibitem{Heil2005}M. Heil, S. Dababneh, A. Juseviciute, F. K{\"a}ppeler, R. Plag, R. Reifarth and S. O'Brien, Phys. Rev. C. \textbf{71}, 025803 (2005).
\bibitem{Reif2008}R. Reifarth \textit{et al.}, Phys. Rev. C, \textbf{77},  015804 (2008).
\bibitem{Beer1992}H. Beer, F. Voss and R. R. Winters, Astrophys. J., Suppl. Ser. \textbf{80}, 403--424 (1992).
\bibitem{PINO}R. Reifarth, M. Heil, F. K{\"a}ppeler and R. Plag, Nucl. Instrum. Method A, \textbf{608}, 139--143 (2009).
\bibitem{simlit}M. Friedman \textit{et al.}, Nucl. Instrum. Method. A, \textbf{698}, 117--126 (2013).
\bibitem{Herrera2015}M. Errera, G. A. Moreno and A. J. Kreiner, Nucl. Instrum. Method. B, \textbf{349}, 64--71 (2015).
\bibitem{Pachuau2017}R. Pachuau, B. Lalremruata, N. Otuka, L. R. Hlondo, L. R. M. Punte and H. H.  Thanga, Nucl. Sci. Eng., \textbf{187}, 70--80 (2017).
\bibitem{LiLiTreview}M. Paul \textit{et al.}, Eur. Phys. J. A, \textbf{55}, 44 (2019).
\bibitem{FRANZ}K. Sonnabend \textit{et al.}, J. Phys. Conf. Ser., \textbf{665}, 012022 (2016).
\bibitem{RooUnfold}T. Adye, Unfolding algorithms and tests using RooUnfold, in: PHYSTAT 2011 Workshop on Statistical Issues Related to Discovery Claims in Search Experiments and Unfolding, CERN, Geneva, Jan (2011). arXiv:1105.1160.
\bibitem{bayesUnfold}G. D'Agostini, Nucl. Instrum. Method. A, \textbf{362}, 487--498 (1995).
\bibitem{Iliadis2011}C. Illadis \textit{et al.}, Astrophys. J. Suppl. S. 193 16 (2011).
\bibitem{Diehl2006}R. Diel \textit{et al.}, Nature 439 7072 45--47 (2006).
\bibitem{Trautvetter86}H. P. Trautvetter \textit{et al.}, Z. Phys. A 323 (1) 1–11 (1986).
\bibitem{Koehler97}P. Koehler, R. Kavanagh, R. Vogelaar, Y. Gledenov and Y. Popov, Phys Rev. C 56 (2) 1138–1143 (1997).
\bibitem{DeSmet07}L. De Smet, C. Wagemans, J. Wagemans, J. Heyse and J. Van Gils, Phys. Rev. C 76  045804 (2007).
\bibitem{TENDL}A.J. Koning and D. Rochman, Nucl. Data Sheets 113 2841 (2012). 
\bibitem{Gastis16}P. Gastis \textit{et al.}, in: Proc. 14th Int. Symp. on Nuclei in the Cosmos (NIC2016).
\bibitem{Kuvin19}S. Kuvin, \textit{et al}. "Constraining the $\nu p$--process through the study of neutron-induced charged-particle reactions on short-lived $^{56}$Ni." Bulletin of the American Physical Society 64 (2019).
\bibitem{Rapp06}W. Rapp, J. G{\"o}rres, M. Wiescher and H. Schatz and F. K{\"a}ppeler, Astrophys. J. 653 474 (2006).
\bibitem{LASI}G. Berg, Design of a Large Angle Spectrometer for Measurements of Protons at 3--7 MeV, Tech. Report, Notre Dame Univ. IN, 18/9/2018. 
\bibitem{Abel19}E. P. Abel, et al., J. Phys. G 46 100501 (2019).

\end{thebibliography}

\end{document}